\documentclass[aps,preprint]{revtex4}%
\usepackage{amsfonts}
\usepackage{amsmath}
\usepackage{amssymb}
\usepackage{graphicx}%
\begin{document}
\title{Blackbody Radiation, Conformal Symmetry, and the Mismatch Between Classical
Mechanics and Electromagnetism}
\author{Timothy H. Boyer}
\affiliation{Department of Physics, City College of the City University of New York, New
York, NY 10031}

\begin{abstract}
The blackbody radiation problem within classical physics is reviewed. \ It is
again suggested that conformal symmetry is the crucial unrecognized aspect,
and that only scattering by classical electromagnetic systems will provide
equilibrium at the Planck spectrum. \ It is pointed out that the several
calculations of radiation scattering using nonlinear mechanical systems do not
preserve the Boltzmann distribution under adiabatic change of a parameter, and
this fact seems at variance with our expectations in connection with
derivations of Wien's displacement theorem. \ By contrast, the striking
properties of charged particle motion in a Coulomb potential or in a uniform
magnetic field suggest the possibility that these systems will fit with
classical thermal radiation. \ It may be possible to give a full scattering
calculation in the case of cyclotron motion in order to provide the needed
test of the connection between conformal symmetry and classical thermal radiation.

\end{abstract}
\maketitle

\subsection{Introduction}

The blackbody radiation problem remains unsolved within classical physics.
\ Although the introduction of energy quanta a century ago has led to the
currently accepted explanation within quantum theory, there is still no firm
conclusion as to whether or not blackbody radiation can be explained within
classical physics. \ In this article we discuss the current situation and
introduce new arguments which again\cite{Boyer1989} suggest that the classical
solution requires the restriction to purely electromagnetic systems where
conformal symmetry is involved.

A century ago, the mismatch between mechanics and electromagnetism was clearly
evident. \ Traditional classical mechanics is invariant under Galilean
symmetry transformation whereas Maxwell's equations are invariant under
Lorentz transformation. \ Also, traditional classical mechanics contains no
scales or fundamental constants, whereas classical electromagnetism contains
several fundamental constants, including a limiting speed of light in vacuum
$c,$ a smallest electronic charge $e$, and Stefan's blackbody radiation
constant\ $a_{S}$. \ It follows that classical mechanics allows separate
scalings of length, time, and energy, whereas classical electromagnetism
allows only a single scaling which couples together the scales of length,
time, and energy in conformal symmetry. \ Although the mismatch between
mechanics and electromagnetism in connection with relativity has been decided
in favor of the Lorentz transformation of electromagnetism being the more
fundamental, there has been no such consensus regarding scaling. \ In the
twentieth century, the mismatch between mechanics and electromagnetism led to
the development of a new mechanics, quantum mechanics, which ties together the
scaling of time and energy through Planck's constant $\hbar$ while still
allowing any mechanical potential to enter the theory. \ Here we again suggest
the alternative resolution to the mismatch which regards classical
electromagnetism as the more fundamental theory. \ Perhaps the restriction to
"mechanical" systems which appear from electromagnetic sources holds the key
to a classical understanding of blackbody radiation and also of at least some
parts of atomic theory. \ This suggestion has recently been bolstered by Cole
and Zou's simulation work obtaining the hydrogen ground state within classical
physics.\cite{Cole&Zou}

In order to describe as much of nature as possible, classical electromagnetic
theory must include classical electromagnetic zero-point radiation as the
homogeneous boundary condition for Maxwell's equations.\cite{Boyer1975}
\ Zero-point radiation is random radiation which is homogeneous in space,
isotropic in direction, and invariant under Lorentz transformation.
\ Furthermore, it turns out that the zero-point radiation spectrum is the
unique spectrum invariant under conformal transformation.\cite{Boyer1989b}
\ The invariance requirements determine the spectrum up to a multiplicative
constant as an energy $\mathcal{U}$ per normal mode given by $\mathcal{U}%
=const\times\omega$ where $\omega$ is the angular frequency of the radiation
mode. \ In order to reproduce the experimentally observed van der Waals
forces, the constant must be chosen as approximately $const=0.525\times
10^{-34}J\sec$, recognizable as $const=(1/2)\hbar$ where $\hbar$ has the
magnitude of Planck's constant. \ 

Now if random classical zero-point radiation is present in the classical
electromagnetic theory at zero temperature, then all the ideas of tradition
classical statistical mechanics are invalid, except as high-temperature
limits. (Just such a situation is also found in quantum theory.) \ Also, if
zero-point radiation is present, it is possible to give derivations of the
Planck spectrum of thermal radiation using classical physics from several
different points of view: energy equipartition for translational degrees of
freedom\cite{Boyer1969}, thermal fluctuations above zero-point
radiation\cite{Boyer1969b}, comparisons between diamagnetic and paramagnetic
behavior\cite{Boyer1983}, the acceleration of point electromagnetic systems
through zero-point radiation\cite{Boyer1984}, and maximum entropy ideas
connected with Casimir forces\cite{Boyer2003b}.

However, despite these derivations, the problem of classical radiation
equilibrium is not completely solved by the introduction of classical
electromagnetic zero-point radiation because the Planck spectrum with
zero-point radiation can be shown to be unstable under scattering by charged
nonlinear mechanical systems. \ In 1924 van Vleck\cite{VanVleck} solved the
Fokker-Planck equation for a general class of charged nonlinear mechanical
systems in random radiation (in the electric dipole approximation and
small-charge limit) with the conclusion that the mechanical system achieved
equilibrium with random radiation only when the mechanical system was
distributed according to the Boltzmann distribution and the radiation
corresponded to the Rayleigh-Jeans spectrum. \ There were further
calculations\cite{Boyer1976}\cite{Boyer1978} coming to this same conclusion in
the 1970's and 1980's, including heroic calculations by Blanco, \ et
al.\cite{Blanco} regarding a relativistic charged particle in certain classes
of mechanical potentials.

These scattering calculations would seem to settle the matter in the negative
from the most fundamental point of view. \ After all, radiation equilibrium
means the stability of the radiation spectrum under scattering by a charged
particle. \ However, there is just one failure of the scattering
calculations--all of them excluded scattering by a Coulomb potential or indeed
by any purely electromagnetic system.

Here we are suggesting that this exclusion contains the very essence for
understanding classical radiation equilibrium \ All of the scattering
calculations to date involve potentials which contradict the conformal-related
scaling symmetries of thermal radiation. \ We suggest that only when purely
electromagnetic scattering systems are used can we hope to obtain the thermal
radiation spectrum observed in nature.

\subsection{Outline of the article}

The outline of our discussion is as follows. \ We begin by noting the
restrictive form of scaling symmetry which follows from conformal symmetry and
which holds in electromagnetism. \ We then confirm the validity of this
scaling for thermal radiation. \ Next we discuss the contrasting
nonrelativistic scattering calculations which appear in the literature. \ It
is pointed out that the nonrelativistic nonlinear scattering systems do not
retain a thermodynamic distribution under an adiabatic compression. \ Thus
they do not allow a derivation of the Wien displacement law. \ We suggest that
such systems merely illustrate the mismatch between classical mechanics and
electromagnetism. \ When we turn to two purely electromagnetic systems in
thermal radiation at $T=0,$ in zero-point radiation, we find that both systems
have very special relativistic properties which suggest possibilities for
thermal equilibrium which are are lacking in other mechanical scattering
systems. \ Finally we give a closing summary.

\subsection{Discussion of Fundamental Constants}

Nonrelativistic classical mechanics has no fundamental constants. \ Thus there
is no preferred length, time, or energy, nor any fundamental connection among
them. \ Accordingly we may choose independent scales of length, of time, and
of energy, and indeed the commonly-used systems of units reflect this
independence. \ Thus given any nonrelativistic mechanical system, a second
system may be constructed which has twice the spatial dimensions, three times
\ the speed, and four times as much energy. \ In contrast, nature associates
three fundamental constants with the observed solutions of Maxwell's
electromagnetic theory. \ These are the largest speed of electromagnetic waves
$c$, the smallest amount of charge $e$, and Stefan's constant $a_{S}$ of
blackbody radiation. \ Each of these constants is left unchanged under
transformations of the conformal group, the largest group of transformations
which leaves invariant Maxwell's equations. \ 

The three constants $c$, $e$, and $a_{S}$ connect together the scales of
length, time, and energy which enter electromagnetism. \ Thus the fundamental
speed $c$ joins the scales of length and time. \ The elementary charge $e$
couples the scales of energy and distance through the potential energy $U$ of
two elementary charges separated by a distance $r$, $U=e^{2}/r$. \ Stefan's
constant $a_{S}$ couples energy density $u$ to temperature $T$; however, since
the temperature scale is not fundamental, we may consider Stefan's constant
$a_{S}$ divided by Boltzmann's constant $k_{B}$ to the fourth power so that
Stefan's law connecting total thermal energy density $u$ to the energy
$k_{B}T$ of long-wavelength modes can be rewritten as $u=(a_{S}/k_{B}%
^{4})(k_{B}T)^{4}$. \ Evidently Stefan's constant (in the form $a_{S}%
/k_{B}^{4}$) again connects the energy scale and the length scale. \ Indeed,
since Stefan's constant can be reexpressed\cite{Morse} in terms of Planck's
constant $\hbar$, \ $a_{S}/k_{B}^{4}=$ $\pi^{2}/(15\hbar^{3}c^{3})$, we could
just as well have chosen $c,$ $e$, and $\hbar$ as our fundamental constants of
electromagnetic theory rather than $c$, $e$, and $a_{S}$. \ We have chosen to
start with Stefan's constant $a_{S}$, which was introduced in 1879, rather
than Planck's constant $h=2\pi\hbar$, which was introduced in 1900, so as to
emphasize that the fundamental constants arise in connection with solutions of
classical electromagnetic theory and need not have any connection with ideas
of energy quanta.

Maxwell's equations are conformal invariant\cite{Cunningham}, retaining
exactly the same form under a conformal transformation; the solutions of
Maxwell's equations are conformal covariant in the sense that under conformal
transformation one solution of Maxwell's equations is mapped into another
solution of Maxwell's equations. \ Rather than working with the full conformal
group, it is convenient here to consider only one small part of this group,
the dilatations. \ (Invariance under dilatations and Lorentz transformation
implies conformal invariance.) \ Dilatations are the mappings where all
lengths, times, and energies are multiplied by a constant $\sigma$,
$\mathbf{r}\rightarrow$ $\mathbf{r}^{\prime}=\sigma\mathbf{r}$, $t$
$\rightarrow$ $t^{\prime}=\sigma t$, and $U$ $\rightarrow$ $U^{\prime
}=(1/\sigma)U$. \ This arrangement preserves the fundamental electromagnetic
constants-- $c$ with units of length divided by time $r/t=\sigma r/(\sigma
t)=r^{\prime}/t^{\prime}$, $e$ with units of the square root of energy times
length $(Ur)^{1/2}=(U/\sigma)^{1/2}(\sigma r)^{1/2}=(U^{\prime}r^{\prime
})^{1/2}$, and $\hbar$ with units of energy times time $Ut=(U/\sigma)\sigma
t=U^{\prime}t^{\prime}$. \ Since the scaling couples together length, time,
and energy, we will term this "$\sigma_{ltE^{-1}}$ scale-invariance."

We note that nature's solutions to the homogeneous Maxwell's equations couple
the scales of length, time, and energy but contain no special length, time, or
energy. \ Thus an electromagnetic plane wave in vacuum of frequency $\nu$, and
wavelength $\lambda$, and electric field amplitude $E_{0}$ can also be
reinterpreted by an observer using a dilated scale of length, time, and energy
as a plane wave of frequency $\nu^{\prime}=(1/\sigma)\nu$, wavelength
$\lambda^{\prime}=\sigma\lambda$, and amplitude $E_{0}^{\prime}=(1/\sigma
^{2})E_{0}$. \ Similarly, blackbody radiation at temperature $k_{B}T$ in a box
of volume $V$, with total thermal energy $U=(a_{S}/k_{B}^{4})V(k_{B}T)^{4}$
and entropy $S=(4/3)(a_{S}/k_{B}^{4})V(k_{B}T)^{3}$ would be reinterpreted by
an observer using a dilated scale of length, time, and energy as blackbody
radiation at temperature $k_{B}T^{\prime}=(1/\sigma)k_{B}T$, in a volume
$V^{\prime}=(\sigma^{3})V$, with total energy $U^{\prime}=a_{S}V^{\prime
}(k_{B}T^{\prime})^{4}=a_{S}(\sigma^{3}V)(k_{B}T/\sigma)^{4}=U/\sigma$ and
total entropy $S^{\prime}=(4/3)a_{S}V^{\prime}(k_{B}T^{\prime})^{3}%
=(4/3)a_{S}(\sigma^{3}V)(k_{B}T/\sigma)^{3}=S$. \ Thus the entropy of thermal
radiation is unchanged by any $\sigma_{ltE^{-1}}$ scaling transformation.

The $\sigma_{ltE^{-1}}$ scale-invariance of electromagnetism can be continued
from Maxwell's equations over to classical electron theory with point masses,
provided mass $m$ is scaled as $m$ $\rightarrow$ $m^{\prime}=m/\sigma,$
corresponding to the scaling for energy $U=mc^{2}$. \ Indeed Haantjes has
shown\cite{Haantjes} that the conformal invariance of electromagnetism can be
extended to classical electron theory provided we transform a point mass as
$m$ \ $\rightarrow$ \ $m^{\prime}=m/\sigma(x)$ where $\sigma(x)$ is the
space-time dependent scale factor of the conformal transformation. \ 

Maxwell's equations, and indeed thermal radiation satisfy $\sigma_{ltE^{-1}}$
symmetry where the scale factor $\sigma_{ltE^{-1}}$ is an arbitrary real
number, $0<\sigma_{ltE^{-1}}<\infty$. \ If we consider a change of units
corresponding to $\sigma_{ltE^{-1}}$ scaling, then any parameter (such as
volume) which changes under $\sigma_{ltE^{-1}}$ must also have a continuous
range of values $0$ to $\infty$, and is appropriate for adiabatic change in
mechanics and thermodynamics, except in the case of mass where we usually
think in terms of substitution rather than of continuous change of a
parameter. \ Indeed, the scaling $m$ $\rightarrow$ $m^{\prime}=m/\sigma
$\ implies that all masses are available in the theory $0<m<\infty$, and that
there is no special value for mass. \ This aspect is not observed in nature
(for example, the mass of the electron is indeed special) and is regarded here
as an aspect beyond our present electromagnetic considerations.

\subsection{Linear and Nonlinear Oscillator Scatterers}

Around the year 1900, Planck\cite{Planck} considered a very small charged
harmonic oscillator in interaction with random electromagnetic radiation.
\ The random classical radiation is expressed as a superposition of plane
waves with random phases, extending throughout space. \ The radiation provides
a stationary random process for the electric field at any spatial point, and
any scattering system will be driven into random oscillation with an amplitude
of motion which is again a stationary random process. \ Planck showed that in
the small-charge limit, a small linear oscillator (electric dipole
approximation for the radiation interaction) had the same average energy as a
radiation mode of the same frequency. \ Since the harmonic oscillator treated
in the electric dipole approximation does not scatter the radiation into any
second frequency, it comes to equilibrium with \textit{any} isotropic spectrum
of random radiation. Clearly, the small harmonic oscillator does not determine
the equilibrium spectrum of thermal radiation based upon its scattering. Now
if classical Boltzmann statistical mechanics is applied to the oscillator,
then one finds energy equipartition for the oscillator and the Rayleigh-Jeans
spectrum for the radiation with which the oscillator is in equilibrium.
\ However, we have noted that Boltzmann statistical mechanics can not be valid
in any classical system which includes the classical electromagnetic
zero-point radiation which is present in nature. \ Thus the derivation of the
Rayleigh-Jeans spectrum from classical Boltzmann statistical mechanics is
irrelevant to a description of nature, except as a high-temperature limit.

Of far more interest are derivations of radiation equilibrium from nonlinear
scattering calculations. \ A nonlinear oscillator which exchanges energy with
several frequencies will indeed enforce a radiation equilibrium; in general,
it will absorb net radiation energy at one frequency and emit net radiation
energy at a different frequency. \ The energy of the oscillator is balanced,
but for a general radiation spectrum the random radiation is not in
equilibrium since there is a continual transfer of energy from one frequency
to another. \ A scattering calculation for a small nonlinear
oscillator\cite{Boyer1976} shows that this scatterer pushes a general
radiation spectrum toward the Rayleigh-Jeans spectrum. \ This fits with van
Vleck's work\cite{VanVleck} of 1924 using a Fokker-Planck equation for the
behavior of a general class of nonlinear oscillators. \ Van Vleck showed that
nonrelativistic nonlinear oscillators treated in the dipole approximation and
small-charge limit come to equilibrium at the Boltzmann distribution for the
oscillators and the Rayleigh-Jeans spectrum for radiation. \ These
calculations are made using nonrelativistic physics and so include the
possibility of particle velocities exceeding the speed of light $c$; however,
since the electric dipole approximation is used for the interaction with
radiation, only the frequencies of the motion and not the velocities are
relevant for the interaction with radiation, and hence no contradiction with
relativity becomes apparent. \ Indeed, one can consider relativistic mechanics
for the particle motion in a general class of potentials, and, as Blanco
\textit{et al.} showed\cite{Blanco}, one still arrives at basically the same
uncomfortable conclusion involving a balance for Boltzmann statistics and the
Rayleigh-Jeans spectrum.

What the class of potentials considered in the nonlinear scattering
calculations do not satisfy is conformal symmetry. \ Conformal symmetry
suggests a tight connection between frequency, energy, and spatial extent, and
this tight connection is relevant to the interaction with radiation. \ The use
of relativistic particle mechanics is indeed sufficient to guarantee that the
particle speed does not exceed the speed of light, but this does not come
close to the restrictions of conformal symmetry. \ For example, the nonlinear
oscillator considered in the scattering calculations\cite{Boyer1976}%
\cite{Boyer1978} of 1976 and 1978 involves the same mechanical system treated
by Born.\cite{Born} \ Solving a Hamiltonian
\begin{equation}
H=p^{2}/(2m)+m\omega_{0}^{2}x^{2}/2+\Gamma x^{3}/3
\end{equation}
\ with a perturbative solution\ \
\begin{equation}
x=D_{1}\sin w+D_{2}(3+\cos2w)
\end{equation}
where the amplitudes $D_{1}$ and $D_{2}$ can be written in terms of the action
variable $J$ as
\begin{equation}
D_{1}=[2J/(m\omega_{0})]^{1/2}\text{ \ and }\ \ D_{2}=-\Gamma\omega
_{0}J/(3\omega_{0}^{4}m^{2})
\end{equation}
The hamiltonian and oscillation frequency can also be rewritten as\ \
\begin{equation}
W=J\omega_{0}-5\Gamma^{2}(\omega_{0}J)^{2}/(12\omega_{0}m^{3})\text{ \ \ and
\ \ }\omega=\omega_{0}-5\Gamma^{2}J/(6\omega_{0}^{4}m^{3})
\end{equation}
The strength of the nonlinearity $\Gamma$ determines the ratio of the
amplitudes $D_{1}$ and $D_{2}$ of the first and second harmonics which
determines whether a little or a lot of radiation is exchanged between
$\omega$ and $2\omega$ going into the electric dipole radiation mode labeled
by $l=1$, $m=0.$ \ Thus the ratio of the radiation energy absorbed and emitted
at the fundamental and its second harmonic is freely adjustable through the
arbitrary nonlinear coupling constant $\Gamma.$ \ Such arbitrariness does not
exist for electromagnetic systems satisfying conformal symmetry.

\subsection{Wien Displacement Law and the Mismatch with the Boltzmann
Distribution}

In addition to the arbitrariness in the radiation connection for nonlinear
oscillators, there is a second troubling aspect in connection with the Wien
displacement law. \ Thermodynamics within the context of classical theory
leads to the Stefan-Boltzmann law $u=a_{S}T^{4}$ and to Wien's displacement
law $\mathcal{U}(\omega,T)=\omega f(\omega/T),$ where $u$ is the thermal
energy per unit volume, $a_{S}$\ is Stefan's constant, $\mathcal{U}(\omega,T)$
is the energy per normal mode of (angular) frequency $\omega$ and temperature
$T$, and $f(\omega/T)$ is an unknown function. \ Both these laws are
experimentally observed to hold in nature. \ The derivation of Wien's
displacement law depends upon carrying out a quasi-static change in the
system\cite{Boyer2003}, usually an adiabatic compression of the radiation
which maintains the radiation as a thermal spectrum at a smoothly changing
temperature. \ The reflection of the radiation from a slowly-moving reflecting
surface on a piston is one method of carrying out the adiabatic
compression.\cite{Richtmyer}

Now radiation inside a reflecting-walled cavity can not bring itself to the
thermal equilibrium spectrum. \ Rather, there must be some scattering system
which changes the spectrum of radiation. \ As noted above, using small
nonlinear oscillator scattering systems, we find that equilibrium occurs for
the Boltzmann distribution for the mechanical system and the Rayleigh-Jeans
law for the radiation spectrum. \ Following in the spirit of the traditional
derivation of the Wien displacement theorem, it is interesting to consider an
adiabatic change where both the radiation and the scattering systems are
changed. \ We can consider an ensemble of many identical nonlinear mechanical
scattering systems in thermal equilibrium with random radiation. \ The
radiation acts as a heat bath for the ensemble of nonlinear oscillators. \ If
we now regard the mechanical scattering systems as decoupled from the
radiation, we can carry out adiabatic changes separately for the mechanical
oscillators and for the radiation. \ Now adiabatic compression for thermal
radiation (when the shape of the container is unchanged) is equivalent to a
$\sigma_{ltE^{-1}}$ scale change. \ Thus any spectrum of radiation with an
energy $\mathcal{U}(\omega,T)=\omega f(\omega/T)$ for a mode of frequency
$\omega$ is carried into%
\begin{equation}
\mathcal{U}^{\prime}(\omega^{\prime},T^{\prime})=\omega^{\prime}%
f(\omega^{\prime}/T^{\prime})=\frac{\omega}{\sigma}f\left(  \frac
{\omega/\sigma}{T/\sigma}\right)  =\frac{1}{\sigma}\omega f\left(
\frac{\omega}{T}\right)  =\frac{1}{\sigma}\mathcal{U}(\omega,T)
\end{equation}
and specifically, the Rayleigh-Jeans spectrum at temperature $T$ is carried
into the Rayleigh-Jeans spectrum at $T^{\prime}=T/\sigma$. \ On the other
hand, mechanics tells us that under a change of parameter of a mechanical
system, the action variables $J$ do not change.\cite{Goldstein-J} \ Thus the
probability distribution $P(J,T,b)$ for the action variables of the mechanical
ensemble at temperature $T$ and parameter $b$ does not change when the
parameter $b$ is slowly changed to $b^{\prime}$. \textit{Linear} scattering
systems with no harmonics $H=J\omega$ (which were used in the suggestive
classical derivations\cite{Boyer1969}\cite{Boyer1969b}\cite{Boyer1983}%
\cite{Boyer1984}\cite{Boyer2003b} of the Planck spectrum) are indeed
transformed into distributions which are again in equilibrium with the thermal
radiation spectrum at some new temperature $T^{\prime}$. Indeed
Cole\cite{Cole} has used this behavior of linear oscillator systems to give a
derivation of the Wien displacement theorem. \ Thus for linear oscillators
$H=J\omega,$
\begin{equation}
P(J,T,\omega)=const\exp\left[  -\frac{H}{\omega f(\omega/T)}\right]
=const\exp\left[  -\frac{J}{f(\omega/T)}\right]
\end{equation}
becomes
\begin{equation}
P(J,T,\omega)=const\exp\left[  -\frac{J}{f(\omega/T)}\right]  =const\exp
\left[  -\frac{J}{f(\omega^{\prime}/T^{\prime})}\right]  =P(J,T^{\prime
},\omega^{\prime})
\end{equation}
\ The linear oscillator keeps a thermal distribution but at a new frequency
and new temperature, $\omega/T=\omega^{\prime}/T^{\prime}$. \ However, for
nonlinear oscillators, an adiabatic change in some mechanical parameter takes
the ensemble of mechanical systems away from the Boltzmann distribution.
\ Thus for Born's nonlinear oscillator mentioned above, a change in the
parameter $\Gamma$ does not preserve a Boltzmann distribution. \ There is no
choice of temperature $T^{\prime}$ for which
\begin{equation}
P(J,T,\Gamma)=const\exp\left[  -\frac{H}{k_{B}T}\right]  =const\exp\left[
-\frac{J\omega_{0}-\{5\Gamma^{2}(\omega_{0}J)^{2}/(12\omega_{0}m^{3})\}}%
{k_{B}T}\right]
\end{equation}
equals
\begin{equation}
P(J,T^{\prime},\Gamma^{\prime})=const\exp\left[  -\frac{\{J\omega_{0}%
-5\Gamma^{\prime2}(\omega_{0}J)^{2}/(12\omega_{0}m^{3})\}}{k_{B}T^{\prime}%
}\right]
\end{equation}
for all $J$ if $\Gamma\neq\Gamma^{\prime}$. \ After the adiabatic mechanical
transformation, the nonlinear oscillators are no longer in equilibrium with
the Rayleigh-Jeans spectrum at any new temperature. \ It seems surprising
indeed that the scattering system which is supposed to bring radiation to
equilibrium can not maintain the equilibrium under any adiabatic change.
\ This suggests that these mechanical systems may not be allowed systems in
classical radiation physics. \ It is symptomatic of the mismatch between
mechanics and electromagnetism.\cite{Szilard}

\subsection{Adiabatic Changes and Zero-Point Radiation}

We have suggested that mechanical systems which do not satisfy conformal
symmetry are not suitable for discussing classical radiation equilibrium. \ We
have seen that they involve excessive freedom in their connections with
radiation and also do not behave appropriately under adiabatic changes of
parameters. \ At this point we need to show that there are indeed mechanical
scattering systems for radiation which overcome these objections. \ In this
section, we will limit our attention to temperature $T=0$ where only
zero-point radiation is present. \ We have already remarked that zero-point
radiation is the unique spectrum of random radiation which is invariant under
conformal transformation.\cite{Boyer1989b} \ We suggest that purely
electromagnetic scattering systems (which are related to conformal symmetry)
will not scatter zero-point radiation toward a new spectrum but will give
radiation equilibrium at temperature $T=0$. \ 

We emphasize that allowed systems should not exchange energy with zero-point
radiation during adiabatic changes. \ Now the zero-point radiation is
invariant under any adiabatic change. \ However, when a mechanical parameter
is changed adiabatically, the mechanical system takes on a new average energy
and a new frequency pattern. \ The mechanical system and radiation (in the
small-charge limit) can be regarded as two separate thermodynamic systems
which can be brought into contact through the electric charge. An average
exchange of energy between the mechanical system and the radiation during an
adiabatic change suggests a change in entropy, and at $T=0$ the ideas of
thermodynamics suggest that no changes of entropy are possible.

Some aspects of this problem were explored\cite{Boyer1978b} in 1978. \ It was
pointed out that all the small nonrelativistic mechanical systems without
harmonics behaved appropriately under changes of mechanical parameters in
zero-point radiation. \ In zero-point radiation the distribution of action
variables for these systems takes the form
\begin{equation}
P(J)=\frac{1}{\hbar/2}\exp\left[  -\frac{J}{\hbar/2}\right]
\end{equation}
and has no dependence upon any mechanical parameters. \ Such systems include
point harmonic oscillator systems in several dimensions and in magnetic
fields, and also nonrelativistic cyclotron motion for a charge in a magnetic
field. \ The scattering systems described in the earlier work are treated in
the electric dipole approximation and interact with radiation at single
frequencies without coupling to any harmonics. \ Thus there is no exchange of
radiation between radiation modes of different frequency, and hence no
radiation equilibrium is forced by the mechanical scattering systems. \ 

However, this is a limiting approximation made for small systems when the
speed of the particle is close to zero. \ Any finite-velocity motion by a
charged particle entails the emission of radiation at all the harmonics of the
fundamental frequency with a distribution of radiated energy among the
harmonics which is determined by the parameter $\beta=v/c$. \ For example, a
charged particle $e$ moving in the xy-plane in a circle of radius $r$ with
speed $v=c\beta$ gives a power radiated per unit solid angle at angle $\theta$
from the $z$-axis at the $n$th harmonic at frequency $n\overline{\omega}=nv/r$
in the form\cite{Jackson695}%
\begin{equation}
\frac{dP_{n}}{d\Omega}=\frac{e^{2}\overline{\omega}^{4}r^{2}}{2\pi c^{3}}%
n^{2}\left\{  \left[  \frac{dJ_{n}(n\beta\sin\theta)}{d(n\beta\sin\theta
)}\right]  ^{2}+\frac{\cot^{2}\theta}{\beta^{2}}J_{n}^{2}(n\beta\sin
\theta)\right\}
\end{equation}
The particle can also absorb energy at each harmonic if there is energy
present in the radiation field. \ Thus all classical electromagnetic systems
of finite size interact with many frequencies and hence determine a spectrum
of radiation equilibrium. \ 

It must be emphasized just how different is this finite-size mechanism for
equilibrium from that involved in point nonlinear mechanical oscillators. For
charged nonlinear scatterers treated in the dipole approximation, the
equilibrium is forced by the mechanical system with its connection between
harmonics depending upon some arbitrary nonlinear parameter. \ For Born's
nonlinear oscillator mentioned earlier, $\Gamma$ is the nonlinear parameter.
\ The nonlinear mechanical oscillator contains within itself the ratios of the
amplitudes for the harmonics with no reference to the relative speed
$\beta=v/c$ of the particle and the radiation. On the other hand, the finite
size of purely harmonic motions gives an electromagnetic basis for forcing
equilibrium. \ \ For uniform circular motion, the relative speed $\beta=v/c$
of the particle and the radiation completely determines the relative power
emitted into the various harmonics.

In addition to forcing an equilibrium radiation spectrum, finite-size systems
have new possibilities for their distributions $P(J,\omega_{0})$ of action
variables in zero-point radiation. \ We must determine whether the adiabatic
invariance of the distribution $P(J)$ which held in zero-point radiation in
the nonrelativistic limit continues for the full relativistic treatment.

\subsection{Aspects of Scattering by Relativistic Cyclotron Motion}

\subsubsection{Equations of Motion}

The simplest purely electromagnetic scattering system of which we are aware is
cyclotron motion, the circular motion of a charged particle in a uniform
magnetic field. \ Here we wish to point out some of the aspects of scattering
by this conformally covariant system. \ We hope to complete and report on a
full scattering calculation in the not distant future.

When we ignore the connection to radiation, cyclotron motion of a particle of
charge $e$ and mass $m$ in a uniform magnetic field $\mathbf{B}$ (in the lab
frame) is described by the hamiltonian%
\begin{equation}
H=\sqrt{(\mathbf{p}-\frac{e}{c}\mathbf{A})^{2}c^{2}+m^{2}c^{4}}\text{
\ \ \ where \ \ \ }\mathbf{A}(\mathbf{r},t)=\frac{\mathbf{B}\times\mathbf{r}%
}{2c}%
\end{equation}
The equations of motion follow as
\begin{equation}
m\frac{d}{dt}\left(  \frac{\mathbf{v}}{(1-v^{2}/c^{2})^{1/2}}\right)
=e\frac{\mathbf{v}}{c}\times\mathbf{B}\text{ \ \ \ \ \ \ \ }%
\end{equation}
or
\begin{equation}
\frac{d^{2}\mathbf{r}}{d\tau^{2}}=\frac{d\mathbf{r}}{d\tau}\times
\overrightarrow{\mathbf{\omega}_{0}}\text{ \ \ \ with \ \ \ }\overrightarrow
{\mathbf{\omega}_{0}}=e\mathbf{B}/(mc)
\end{equation}
\ where $\tau$ is the particle proper time
\begin{equation}
d\tau=dt/\gamma=\sqrt{1-v^{2}/c^{2}}dt
\end{equation}
We note from Eq. (14) that (independent of the orbital radius and velocity)
the orbital rotation rate is always $\omega_{0}$ when measured using the
particle's proper time. \ Taking the charge $e$ as positive and the magnetic
field in the negative $z$-direction, the solutions correspond to uniform
circular motion at frequency $\overline{\omega}=\omega_{0}/\gamma$ with
$\gamma=(1-v^{2}/c^{2})^{-1/2},$%
\begin{equation}
x-x_{0}=r\cos[\overline{\omega}t+\phi]\text{ \ \ \ \ \ \ }y-y_{0}%
=r\sin[\overline{\omega}t+\phi]\text{ \ \ \ where \ \ \ }\overline{\omega
}=\omega_{0}/\gamma
\end{equation}
The angular momentum $J$, including both mechanical and electromagnetic field
angular momentum is an adiabatic invariant\cite{Jackkson-J}%
\begin{equation}
J=m\gamma vr-\frac{eBr^{2}}{2c}=\frac{1}{2}m\gamma vr=\frac{eBr^{2}}{2c}%
\end{equation}
where the last two forms follow from the equation of motion for the circular
orbit, $m\gamma v^{2}/r=evB/c$. \ The action variable $J$ determines the orbit
radius and also the orbit velocity as
\begin{equation}
r=\sqrt{\frac{2J}{m\omega_{0}}}\text{ \ \ \ \ }\beta=\sqrt{\frac{2J\omega
_{0}/(mc^{2})}{1+2J\omega_{0}/(mc^{2})}}\text{ \ \ \ \ }\gamma=\sqrt
{1+\frac{2J\omega_{0}}{mc^{2}}}%
\end{equation}
and these expressions hold for all $J$ and $\omega_{0}$. \ As $J$ ranges over
the interval $(0,\infty)$, the velocity parameter $\beta$ (or $\gamma$) is a
monotonically increasing function of $J$ for every choice of $m$ or
$\omega_{0}$. \ There is no preferred value of $\beta$ or $\gamma$.

\subsubsection{Nonrelativistic Limit of Cyclotron Motion}

In the limit of nonrelativistic motion in the lab frame, $2J\omega_{0}%
/(mc^{2})<<1$, cyclotron motion in classical zero-point radiation was
treated\cite{Boyer1978b}\cite{Boyer1980} in 1978 and 1980. \ \ In this limit,
all the motion is at the single frequency $\omega_{0}=eB/(mc)$, since
$\overline{\omega}\rightarrow\omega_{0}$ as $\gamma\rightarrow1$. \ A
Fokker-Planck equation was obtained for the probability distribution $P(J)$ of
the action variable $J$ in a random radiation spectrum with energy per normal
mode $\mathcal{U}(\omega)=(1/2)\hbar\omega$ corresponding to zero-point
radiation. \ The Fokker-Planck equation gave the probability distribution of
Eq. (10). \ This distribution seems exactly appropriate for zero-point
radiation; even when the magnetic field $B$ is changed, and hence the
nonrelativistic frequency $\omega_{0}=eB/(mc)$ is changed, the mechanical
motion remains in equilibrium with the zero-point radiation and does not
exchange any energy (on average) with the zero-point radiation. \ All changes
of average mechanical energy when the magnetic field $B$ is changed are due to
the Faraday-induced electric field associated with the changing $B$.

\subsubsection{Relativistic Treatment of Cyclotron Motion}

However, what happens when we go to the full relativistic treatment? \ In the
relativistic treatment, the mechanical motion varies in frequency with $J$
since $\overline{\omega}=\omega_{0}/\gamma$, and also the charge interacts
with radiation at all the harmonics of the mechanical frequency. \ Do we still
have radiation equilibrium? \ Do we still have $P(J)$ independent of
$\omega_{0}$ in order to maintain our ideas of entropy under adiabatic changes
of magnetic field at temperature $T=0$? \ 

Now relativistic cyclotron motion is an electromagnetic system satisfying
conformal invariance. \ Thus we expect that the conformal-invariant zero-point
radiation spectrum is maintained and that the distribution $P(J)$ maintains
the form given in Eq. (10). \ In order to prove this we need a complete
calculation of the radiation scattering. \ However, for the present we will
present some suggestive evidence. \ The Fokker-Planck equation needed to
obtain $P(J)$ requires calculations of the radiation energy loss per unit time
by the mechanical particle motion, the average energy absorbed per unit time
from zero-point radiation, and the average of the square of the energy
absorbed per unit time from zero-point radiation. \ It is easy to calculate
the radiated energy per unit time for a charged particle $e$ moving in a
circle of radius $r$ with frequency $\overline{\omega}$ as
\begin{equation}
P_{emitted}=\frac{2}{3}\frac{e^{2}}{c^{3}}\overline{\omega}^{4}\gamma^{4}r^{2}%
\end{equation}
where $\gamma=(1-v^{2}/c^{2})^{-1/2}=[1-(r\overline{\omega}/c)^{2}]^{-1/2}$.
\ In the nonrelativistic limit, the radiation emission for cyclotron motion
is
\begin{equation}
P_{emitted}^{cyclotronNR}=\frac{2}{3}\frac{e^{2}}{c^{3}}\omega_{0}^{4}r^{2}%
\end{equation}
since in the nonrelativistic limit $\gamma\rightarrow1$ and $\overline{\omega
}=\omega_{0}/\gamma\rightarrow\omega_{0}$. \ But now notice when we substitute
the fully relativistic expressions $\overline{\omega}=\omega_{0}/\gamma$ and
$r=\sqrt{2J/(m\omega_{0})}$ for cyclotron motion into the fully relativistic
expression for $P_{emitted}$. \ We find
\begin{equation}
P_{emitted}^{cyclotronR}=\frac{2}{3}\frac{e^{2}}{c^{3}}\overline{\omega}%
^{4}\gamma^{4}r^{2}=\frac{2}{3}\frac{e^{2}}{c^{3}}\omega_{0}^{4}\frac
{2J}{m\omega_{0}}\text{\ for all }J\text{ and }\omega_{0}%
\end{equation}
This expression is identical with the nonrelativistic expression. \ When
written in terms of $J$ and $\omega_{0}$, the expression makes no explicit
reference to any velocity and retains its nonrelativistic form for
\textit{all} values of $J$ and $\omega_{0}$. \ If we go to the inertial frame
in which the charge is at rest at some instant, then for a small time interval
the particle motion is nonrelativistic and the charge is found moving in
circular arcs with the same frequency $\omega_{0}=eB/(mc)$ as is involved in
nonrelativistic motion and with the same $J$ as is involved in the lab motion.
\ Also, the zero-point radiation spectrum is Lorentz invariant and hence the
same in any inertial frame. \ Thus in the momentarily comoving reference
frame, where the motion is nonrelativistic, cyclotron motion seems to take the
same form as for nonrelativistic motion in the lab frame. \ This suggests the
possibility that relativistic cyclotron motion will maintain the same
distribution $P(J)$ in Eq. (10) which is invariant under adiabatic changes,
exactly as required for our ideas of thermodynamic equilibirum at $T=0$.

Furthermore, the connection between the orbit and the relative energy radiated
into various harmonics is not at our disposal, as it is in the nonlinear
oscillator case, but rather is tightly connected to formulae involving
spherical Bessel functions. \ This suggests the possibility that this purely
electromagnetic system will allow equilibrium with zero-point radiation.

\subsubsection{Relativistic Limit of the Harmonic Potential}

In order to emphasize that cyclotron motion has very special properties not
encountered with nonelectromagnetic systems, we can consider relativistic
motion in a harmonic oscillator potential in the lab frame, when limiting
ourselves to circular orbits. \ In the nonrelativistic limit, motion in a
harmonic oscillator potential $V_{SHO}(r)=(1/2)kr^{2}$ is at the frequency
$\omega_{0}=\sqrt{k/m}$ and involves no harmonics. \ In terms of relativistic
particle mechanics for a circular orbit in this same potential,%

\begin{equation}
m\gamma\frac{v^{2}}{r}=kr\text{ \ \ \ \ }J=m\gamma vr\text{ \ \ \ \ }%
\end{equation}
Combining these expressions gives%
\begin{equation}
\text{\ }\gamma^{3}\beta^{4}=\left(  \frac{J\omega_{0}}{mc^{2}}\right)
^{2}\text{ \ with \ \ }\omega_{0}=\sqrt{\frac{k}{m}}\text{ \ \ and \ }%
\beta=\frac{\overline{\omega}r}{c}\text{ \ \ \ \ }\overline{\omega}%
=\frac{\omega_{0}}{\gamma^{1/2}}%
\end{equation}
Solving the equation connecting $\beta$ and $J$, we find that
\begin{equation}
\text{\ \ \ \ }\beta\approx\sqrt{\frac{J\omega_{0}}{mc^{2}}}\text{
\ \ \ }r\approx\left(  \frac{J}{m\omega_{0}}\right)  ^{1/2}\text{\ \ \ \ for
}\frac{J\omega_{0}}{mc^{2}}<<1
\end{equation}
and%
\begin{equation}
\text{ }\gamma\approx\left(  \frac{J\omega_{0}}{mc^{2}}\right)  ^{2/3}%
\text{\ \ \ \ }r\approx\left(  \frac{Jc}{m\omega_{0}^{2}}\right)  ^{1/3}\text{
\ \ for }\frac{J\omega_{0}}{mc^{2}}>>1
\end{equation}
Then in terms of the parameter $J\omega_{0}/(mc^{2})$, radiation emission is
given by
\begin{equation}
P_{emission}^{circularSHO}\approx\frac{2}{3}\frac{e^{2}}{c^{3}}(c\omega
_{0})^{2}\left(  \frac{J\omega_{0}}{mc^{2}}\right)  \text{ for }\frac
{J\omega_{0}}{mc^{2}}<<1
\end{equation}
while%
\begin{equation}
P_{emission}^{circularSHO}\approx\frac{2}{3}\frac{e^{2}}{c^{3}}(c\omega
_{0})^{2}\left(  \frac{J\omega_{0}}{mc^{2}}\right)  ^{2}\text{ for }%
\frac{J\omega_{0}}{mc^{2}}>>1
\end{equation}
The change from linear over to quadratic dependence on $J$ shows clearly that
$P_{emission}^{circularSHO}$ does not retain its nonrelativistic functional
form for the harmonic oscillator potential. \ 

Indeed we notice the relation $\overline{\omega}=\omega_{0}/\gamma^{1/2}$
which holds for the harmonic oscillator potential is not connected to particle
proper time. \ If we go to the momentarily comoving reference frame in this
case, the rotation frequency in this frame depends upon the speed of the
particle in its orbit in the lab frame; it does \textit{not} take the
nonrelativistic harmonic oscillator\ value $\omega_{0}=\sqrt{k/m}$. \ Thus in
zero-point radiation, the energy pick up and loss in the instantaneous rest
frame of the particle depends upon a frequency which varies with the velocity
of the particle in the lab frame. \ The pick up and loss of energy in the
momentarily comoving reference frame of the particle do not take the same form
as for nonrelativistic motion. \ Relativistic nonelectromagnetic systems do
not have characteristics suitable for thermodynamic equilibrium with
zero-point radiation.

\subsection{Comments on the Coulomb Potential}

In the previous section, we focused our attention on cyclotron motion because
this seems the simplest electromagnetic scattering system; cyclotron motion
has an easily-calculable nonrelativistic limit for large mass $m$, and in
zero-point radiation seems to retain its nonrelativistic forms at all
velocities when expressed in terms of $J$ and $\omega_{0}$. \ However, despite
its complications, the Coulomb potential allows some interesting
observations.\cite{Cole1989}

The Coulomb potential $V_{C}(r)=e^{2}/r$ is the only potential of the form
$V(r)=k/r^{n}$ where the constant $k$ giving the strength of the potential
does not change under a $\sigma_{ltE^{-1}}$ scale transformation. \ Since an
energy must transform as $1/\sigma$, $V^{\prime}=(1/\sigma)V,$ we have
\begin{equation}
V^{\prime}(r^{\prime})=k^{\prime}/r^{\prime n}=k^{\prime}/(\sigma^{n}%
r^{n})=(k^{\prime}/\sigma^{n})/r^{n}=(1/\sigma)k/r^{n}=(1/\sigma)V(r)
\end{equation}
so that only for $n=1$ do we have $k^{\prime}=k.$ \ Thus the electronic charge
$e$ appearing in the Coulomb potential is invariant under conformal
transformation and no other potential-strength constant is so invariant. \ 

Since for non-Coulomb potentials the constant $k$ changes with the choice of
scale $\sigma_{ltE^{-1}}$, $k$ must be treated as a parameter subject to
variation $0<k<\infty,$ and can be used to carry out adiabatic changes in the
mechanical system. \ Using such adiabatic changes, it may well be possible to
transfer energy from one frequency range to another in the presence of
zero-point radiation, hence violating our ideas of entropy changes at
temperature $T=0$. \ On the other hand for the Coulomb potential, the strength
parameter $e^{2}$ is scale invariant and hence is not subject to adiabatic change.

The hamiltonian for a mass $m$ in the Coulomb potential $H=(p^{2}c^{2}%
+m^{2}c^{4})^{1/2}+e^{2}/r$ can be rewritten in terms of action-angle
variables as\cite{Goldstein498}%
\begin{equation}
H=mc^{2}\left(  1+\frac{(e^{2}/c)^{2}}{[(J_{3}^{\prime}-J_{2}^{\prime}%
)+\sqrt{J_{2}^{\prime2}-(e^{2}/c)^{2}}]^{2}}\right)  ^{-1/2}%
\end{equation}
We note that $H/(mc^{2})$ involves only the action variables $J_{2}^{\prime},$
$J_{3}^{\prime},$ and the quantity $e^{2}/c.$ \ In the presence of zero-point
radiation (which is scale invariant and indeed invariant under conformal
transformation), the only scale for length, time, or energy is through the
mass $m$, and the pattern of velocities is the same independent of the mass
$m.$ \ Thus for a point charge in a Coulomb potential in zero-point radiation,
we can not obtain a nonrelativistic limit by considering a large-mass limit.
\ Rather, if one can find the solution for this classical hydrogen atom for
one choice of the mass $m,$ the same distribution $P(J)$ will hold for any
other mass, while all lengths, times, and energies will be rescaled, and the
velocities will be left unchanged.

Cyclotron motion and Coulomb potential motion in zero-point radiation are very
different. \ Cyclotron motion depends upon the mass $m$ and the pure number
$J\omega_{0}/(mc^{2})=JeB/(m^{2}c^{3})$ where $0<J<\infty.$ \ In the presence
of zero-point radiation, all average quantities depend upon the mass $m$ and
$\hbar\omega_{0}/(mc^{2})$ in the small-charge limit. \ Here there is no
reason for a preferred choice of the value of $\hbar.$ \ However, for the
Coulomb potential, the hamiltonian form (29)\ in terms of action variables
shows that $e^{2}/c$ is a lower bound\cite{Boyer2004} for the action variable
$J_{2}^{\prime}$ so that we require $e^{2}/c<J_{2}^{\prime}<\infty$. \ Now the
values of the action variables in zero-point radiation are dependent upon the
multiplicative constant $\hbar$ giving the scale of the zero-point radiation.
\ Thus this suggests the basis for a connection between $e^{2}/c$ and $\hbar.$
\ If $\hbar$ is too small, then the value of $J_{2}^{\prime}$ will be too
close to the cut-off $e^{2}/c$ which appears in the relativistic mechanics of
the Coulomb potential. \ We again suggest\cite{Boyer1989} that a full
understanding of the behavior of a charged particle in the Coulomb potential
in classical zero-point radiation will lead to a calculation of the fine
structure constant.

\subsection{Closing Summary}

In this work we revisit the suggestion that scattering by classical
electromagnetic systems (which involve conformal symmetry) will provide an
explanation for the Planck spectrum for thermal radiation within the context
of classical physics. \ This time we go beyond the considerations of scaling
symmetry which were mentioned fifteen years ago. \ We suggest that the several
calculations of radiation scattering using nonlinear mechanical systems merely
illustrate the mismatch between mechanics and electromagnetism and are not
relevant for understanding nature.\ \ We point out the curious fact that most
mechanical systems do not preserve the Boltzmann distribution under adiabatic
change of a parameter. \ This fact seems at variance with our expectations in
connection with derivations of Wien's displacement theorem where we expect a
scatterer which enforces an equilibrium spectrum to remain in equilibrium
during a suitable adiabatic change. \ Linear oscillators do not enforce
radiation equilibrium in the nonrelativistic approximation, but indeed do
impose equilibrium when treated relativistically. \ We emphasize some of the
striking properties of charged particle motion in a Coulomb potential or in a
uniform magnetic field which suggest the possibility that these systems will
fit with classical thermal radiation. \ In particular, cyclotron motion
involves linear motion in the nonrelativistic approximation and has surprising
continuities in form when treated relativistically, and the Coulomb potential
is unique in not allowing adiabatic changes of the potential-strength
parameter $e$. \ Finally we note that it may be possible to give a full
scattering calculation in the case of cyclotron motion which should provide a
crucial test of the suggested connection between conformal symmetry and
classical thermal radiation.

Awareness of the mismatch between mechanics and electromagnetism seems to
involve contrasting perspectives between relativistic invariance and conformal
invariance. \ In the last decades of the nineteenth century, physicists became
concerned about the mismatch between mechanics and electromagnetism in
connection with the fundamental constant $c$, the unique value of the speed of
light in vacuum appearing in nature. In the early years of the 20th century,
the relativistic symmetry of Maxwell's equations and its solutions was
recognized, and the constant $c$ was take as a fundamental connection between
the scales of length and time. \ Within classical physics, there has been no
comparable attention to the mismatch between mechanics and electromagnetism
reflected in the fundamental electronic charge $e$ and Stefan's constant
$a_{S}$ (or equivalently Planck's constant $\hbar$) which occur in the
solutions to Maxwell's equations which appear in nature. \ At present these
constants which couple energy and length are not usually associated with the
conformal invariance of Maxwell's equations discovered by Cunningham and
Bateman in 1909.

\subsection{Acknowledgement}

A PSC-CUNY Faculty Research Award was held while this work was completed.


\begin{thebibliography}{99}
\bibitem {Goldstein498}See reference 21, p. 498. \ We are using action
variables which are $1/(2\pi)$ times those in Goldstein's text.
\end{thebibliography}
\end{document}